\documentclass[
superscriptaddress,
 amsmath,amssymb,
 aps,
]{revtex4-2}

\usepackage{xcolor}
\usepackage{graphicx}
\usepackage{dcolumn}
\usepackage{bm}
\usepackage{hyperref}
\usepackage[utf8]{inputenc}
\usepackage[T1]{fontenc}
\usepackage{braket}

\usepackage{physics}
\usepackage{siunitx}
\usepackage{upgreek}
\usepackage[normalem]{ulem}
\begin{document}

\newcommand{\red}[1]{\textcolor{red}{#1}}

\title{Fast, continuous and coherent atom replacement in a neutral atom qubit array}

\author{Yiyi Li}%
\email{These authors contributed equally to this work.}
\affiliation{Department of Electrical and Computer Engineering, Princeton University, Princeton, New Jersey 08544, USA}
\author{Yicheng Bao}%
\email{These authors contributed equally to this work.}
\affiliation{Department of Electrical and Computer Engineering, Princeton University, Princeton, New Jersey 08544, USA}
\author{Michael Peper}%
\email{These authors contributed equally to this work.}
\affiliation{Department of Electrical and Computer Engineering, Princeton University, Princeton, New Jersey 08544, USA}
\author{Chenyuan Li}%
\affiliation{Department of Electrical and Computer Engineering, Princeton University, Princeton, New Jersey 08544, USA}
\affiliation{Department of Physics, Princeton University, Princeton, NJ 08544, USA}
\author{Jeff D. Thompson}
\email{jdthompson@princeton.edu}
\affiliation{Department of Electrical and Computer Engineering, Princeton University, Princeton, New Jersey 08544, USA}

\date{\today}

\begin{abstract}
Neutral atom quantum processors are a promising platform for scalable quantum computing. An obstacle to implementing deep quantum circuits is managing atom loss, which constitutes a significant fraction of all errors. Current approaches are either not capable of replacing lost atoms in the middle of a circuit---and therefore restricted to fixed, short circuit depths---or require more than an order of magnitude longer time than gate and measurement operations to do so. In this work, we demonstrate fast, continuous atom replacement leveraging the metastable $^{171}$Yb qubit. A continuously loaded reservoir near the computation zone enables on-demand atom extraction with tweezers up to 500 times per second. New qubit arrays can be prepared 30 times per second when including single-atom preparation, non-destructive imaging and initialization. Importantly, existing qubits are completely undisturbed by the reloading process, owing to the extreme isolation of the metastable qubit from cooling and imaging light. This work establishes a complete foundation for implementing fast quantum circuits with unlimited depth, removing a final roadblock for fault-tolerant quantum computing with neutral atoms.
\end{abstract}

\maketitle

\section{Introduction}

In recent years, neutral atom quantum computers have made significant advances in scale~\cite{ebadi2021,Scholl2021quantum,manetsch2024tweezer} and gate fidelity~\cite{evered2023,peper2025,Tsai2025Benchmarking}, leading to demonstrations of quantum error correction and algorithms with logical qubits~\cite{bluvstein2022,bluvstein2024logical,reichardt2024fault,bedalov2024fault,zhang2025}. So far, these demonstrations are constrained to a fixed, finite number of circuit layers, limited by the number of qubits initially prepared and the accumulation of leakage and loss through gate errors, atom transport, background gas collisions and, in many cases, destructive qubit readout. Provided these errors can be detected and repaired, the bias towards leakage and loss is in fact, preferable, as it results in lower error correction overhead~\cite{wu2022,ma2023a,scholl2023a,Chow2024Circuit,perrin2025quantum,yu2025processing,baranes2025leveraging}. But in this regime, the timescale to replace atoms can bottleneck the entire computation.

In most current experiments, loading atoms into tweezers is not only the slowest operation, but also destroys existing qubits through the application of magnetic field gradients and resonant cooling and imaging light. In Ref.~\cite{Singh2023mid}, this problem was circumvented using a dual-species atom array, where the coherence of one atomic species is preserved while atoms of the other species are reloaded from an atomic beam. However, the characteristic loading time of \SI{90}{ms} is considerably longer than typical gates and measurements. Other works have demonstrated continuous loading of tweezers from a reservoir on similar timescales without maintaining coherence~\cite{gyger2024continuous,norcia2024iterative}, or mid-circuit measurement without reloading, protecting existing qubits from scattering using shelving~\cite{lis2023a,Graham2023Midcircuit}, light-shifts~\cite{norcia2023a,Hu2025Site} or a distant readout zone~\cite{bluvstein2024logical}.

In this work, we demonstrate fast, continuous mid-circuit reloading of $^{171}$Yb metastable nuclear spin qubits~\cite{ma2023a, lis2023a}. Our approach is based on a dense, continuously fed atom reservoir several hundred microns from the computation zone~\cite{Chen2019Continuous}. From this reservoir, we can repeatedly extract atoms using optical tweezers with a characteristic loading time of only \SI{1}{ms}, and transport them to the computation zone in \SI{1}{ms} of transport time. When repeatedly loading a 256-site tweezer array continuously at a 500 cycles per second, we extract $7.3\times10^{4}$ atoms/s with no observable depletion of the loading rate or reservoir density. We demonstrate repeated preparation and measurement of arrays of singly-occupied tweezers with qubits initialized in the metastable $6s6p\,^3P_0$ state at 30 cycles per second, limited primarily by the time required for imaging, cooling and light-assisted collisions, and achieve repeated measurements at 50 cycles per second by reusing qubits with non-destructive measurements. After transferring to a stationary tweezer array, we show that the lifetime, coherence time and single qubit gate operations on the metastable qubits are completely unaffected by subsequent rounds of reloading.

The mid-circuit tweezer loading time demonstrated in this work is an improvement of nearly two orders of magnitude over the current state of the art. The continuously loaded reservoir is capable of supplying $10^6$ atoms/s given a larger tweezer array and plausible improvements in the speed of light-assisted collisions, imaging and cooling. This work establishes a solid foundation for exploring fully fault-tolerant quantum computation with neutral atom arrays, and will also support zero-deadtime quantum metrology and timekeeping.

\begin{figure*}
    \centering
    \includegraphics[width=\textwidth]{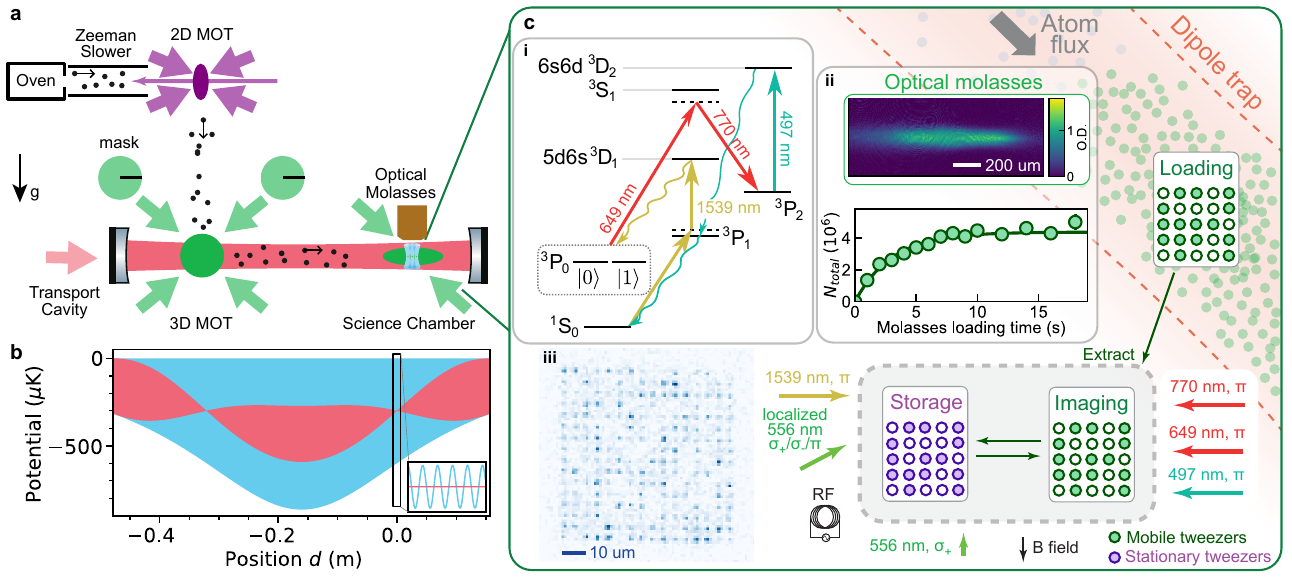}
    \caption{
    \textbf{Continuous reloading approach.}
    (a) Schematic drawing of the apparatus showing laser pre-cooling stages and atom transport to the science chamber in the cavity-enhanced transport ODT.
    (b) The intracavity standing wave in the transport trap (blue) is suppressed by driving with multiple longitudinal modes separated by one free spectral range (red). By tuning the relative amplitude, the lattice can be exactly nulled at the tweezer location (inset). 
    (c) Schematic drawing of the tweezer region, showing the optical molasses, tweezer operation zones (loading, imaging, and storage), and laser beam paths. All of the indicated lasers illuminate all zones, with the exception of the ``localized 556 nm'' path that does not intersect the reservoir.
    (i) Energy levels used to initialize and measure qubits in the metastable $6s6p~^3P_0$ manifold. Optical pumping to $^3P_0$ is performed with a two-photon transition to the $5d6s~^3D_1$ state which decays to $^3P_0$ (yellow). Non-destructive spin readout is performed by sequentially transferring $\ket{0}$ and $\ket{1}$ to $^3P_2$ using coherent Raman transitions (red arrows), followed by depumping through the $6s6d\,^3D_2$ state (blue).
    (ii) Absorption image and loading curve of the reservoir.
    (iii) Single shot image of a 16$\times$16 tweezer array in the imaging zone.
    }
    \label{fig:fig1}
\end{figure*}

\begin{figure*}
    \centering
    \includegraphics[width=0.8\textwidth]{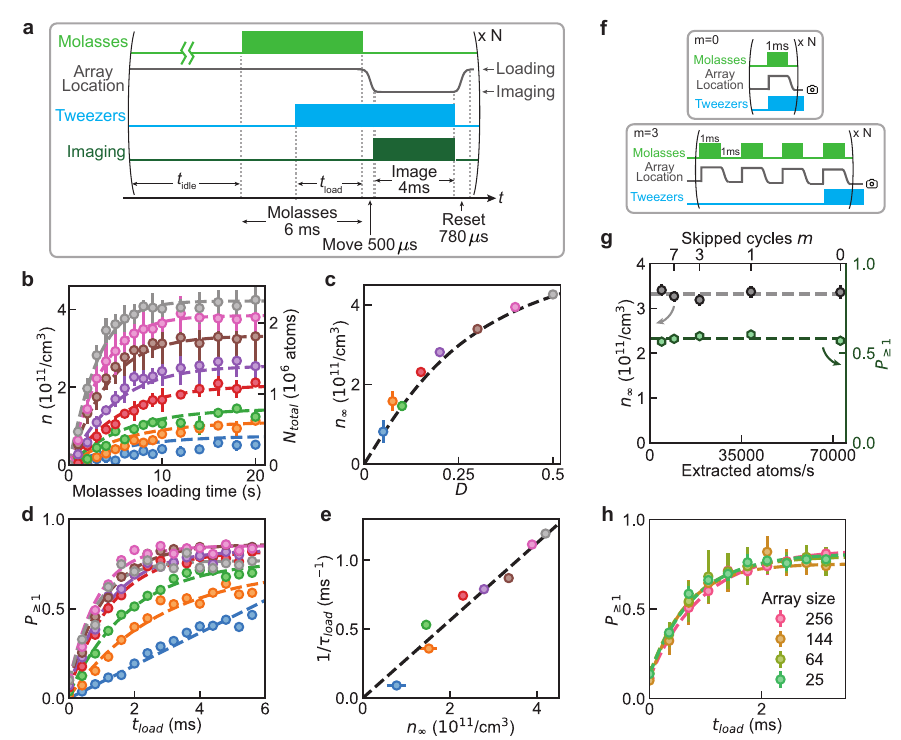}
    \caption{
    \textbf{Fast optical tweezer loading. }
    (a) The tweezers are superimposed with the reservoir for a time $t_{\text{load}}$, then moved to the imaging zone. $t_{\text{idle}}$ is varied to control the molasses duty cycle independently of $t_{\text{load}}$.
    (b) Reservoir loading curves for different molasses duty cycles $D$, showing the reservoir atom number and density measured with absorption imaging. The color of the curves corresponds to the duty cycles indicated in panel (c).
    (c) Steady-state molasses density $n_\infty$ as a function of molasses duty cycle $D$.
    (d) Tweezer loading curves for different molasses duty cycles. The vertical axis shows the probability for a tweezer to have more than one atom, $P_{\geq 1}$. The dashed lines show fits to the function $p_{\infty}(1-e^{-t_{\text{load}}/\tau_{\text{load}}})$, from which the loading time constant $\tau_{\text{load}}$ is extracted.
    (e) $1/\tau_{\mathrm{load}}$ as a function of the steady-state reservoir density. The dashed line is a linear fit with slope 0.28(3)$\times 10^{-11}$~cm$^{3}$/ms.
    (f) To probe the achievable atom extraction rate, we repeatedly load atoms for 1 ms, transport to the imaging zone and release them. The extraction rate is varied by only turning the tweezer on for 1/m cycles. After the last cycle, tweezers are imaged to determine $P_{\geq 1}$.
    (g) Reservoir density $n_\infty$ (gray circles) and $P_{\geq 1}$ as a function of $m$. The reservoir is not measurably depleted by extracting $7.3 \times 10^{4}$ atoms/s.
    (h) Tweezer loading curves with different array sizes ($D=0.5$). The trap depth is maintained at $U_0/h = 1.94$\,MHz, with tweezer spacing of 4.5~$\mu$m.
    }
    \label{fig:fig2}
\end{figure*}

\section{Optical tweezer loading}
The core of our approach is a cold atom reservoir that is continuously loaded from a 3D magneto-optical trap (MOT) in an adjoining vacuum chamber, following the approach of Ref.~\cite{Chen2019Continuous} (Fig.~\ref{fig:fig1}a). Compared to prior work on low latency atom transport with cold lattices~\cite{bao2022,Klostermann2022Fast,norcia2024iterative}, our implementation is optimized for atom flux at the science chamber without regard for the transit time, as the 3D MOT and upstream cooling stages operate continuously. Atoms are guided \SI{35}{cm} from the MOT to the reservoir using an optical dipole trap (ODT) with an in-vacuum enhancement cavity, driven by multiple frequency tones to partially suppress standing waves (Fig.~\ref{fig:fig1}b). Atoms arriving at the reservoir are stopped and cooled using a 1D optical molasses. The loading time constant of the reservoir is approximately 5 seconds, but we operate in the steady-state regime (additional details about the experimental apparatus can be found in App.~\ref{sec:experimentappratus}).

To load optical tweezers from the reservoir, we superimpose a 2D tweezer array with 256 sites generated by cross acousto-optic deflectors (AODs) operating at 488~nm, which is a magic wavelength for the intercombination line~\cite{ma2022}. After loading, the array is moved with a galvanometer to an imaging zone displaced by $\sim\SI{300}{\mu m}$, to avoid fluorescence background from atoms in the reservoir (Fig.~\ref{fig:fig1}c). For the same reason, the molasses beams are switched off during imaging. For short switch-off times,  previously loaded atoms remain in the reservoir, but the average reservoir loading rate is reduced by the duty cycle (Fig.~\ref{fig:fig2}b,c, App.~\ref{sec:loadingdynamics}).

To probe the dependence of the tweezer loading rate on the reservoir density, we use the sequence in Fig.~\ref{fig:fig2}a to scan the tweezer loading time and molasses duty cycle independently. As the duty cycle is varied from 0.05 to 0.5, the steady-state reservoir density ranges from $0.8$ to $4.3\times 10^{11}$~atoms/cm$^3$ (Fig.~\ref{fig:fig2}b,c). The measured tweezer loading time $\tau_{load}$ (defined as the $1/e$ time to load one or more atoms) is inversely proportional to the steady-state density, reaching 0.84~ms for the highest duty cycle (Fig.~\ref{fig:fig2}d,e). The molasses parameters and loading rate are in reasonable agreement with a Monte Carlo simulation of atomic trajectories (App.~\ref{sec:loadingsimulation}).

To probe the maximum rate at which atoms can be extracted from the reservoir, we perform repeated ``grab and drop'' cycles at up to 500~cycles/s with $t_{\text{load}} = \SI{1}{ms}$, without stopping to take images except after the final cycle to verify the loading probability (Fig.~\ref{fig:fig2}f,g). No change to the molasses density and loading probability are observed as the extraction rate is varied from $4.5\times 10^3$~atoms/s to $7.3\times 10^4$~atoms/s. We also observe no change in the loading time constant when varying the array size from 5$\times$5 to 16$\times$16 while keeping the lattice spacing and power per tweezer fixed (Fig.~\ref{fig:fig2}h). At the maximum extraction rate, only 4\% of the 1.6 million~atoms/s reaching the reservoir are loaded into tweezers, which suggests that even higher extraction rates are possible with larger tweezer arrays.

\begin{figure}
    \centering
    \includegraphics[width=0.5 \columnwidth]{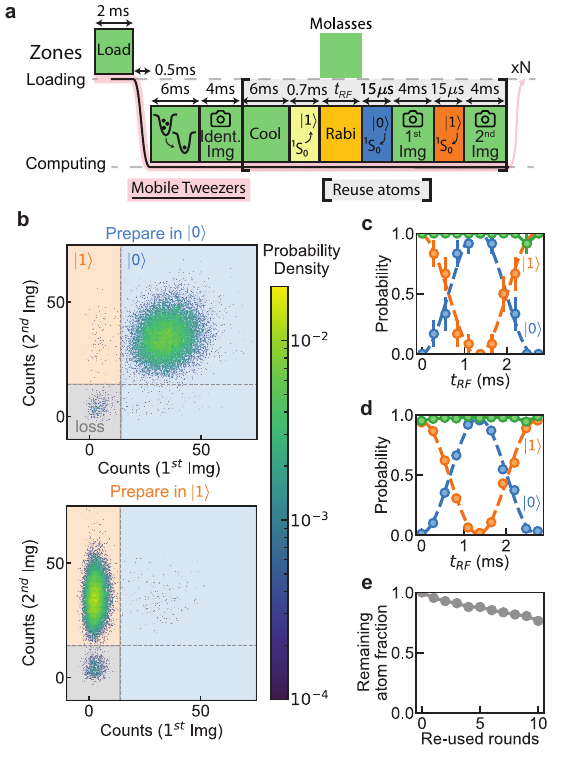}
    \caption{
    \textbf{Loss-resolved spin readout. } 
    (a) Experimental sequence for initialization and readout of the $^3P_0$ qubit (see text). The identification image is used to determine which sites are initially occupied.
    (b) Photon count distributions from the two spin readout images, with the qubit initialized in $\left|0\right\rangle$ or $\left|1\right\rangle$. The shaded regions show the outcome assignment: $\ket{0}$ (blue), $\ket{1}$ (orange), and loss (gray).
    (c)(d) Nuclear spin Rabi oscillation with non-destructive measurement, showing $P_{|0\rangle}$ (blue), $P_{|1\rangle}$ (orange) and $1-P_{\text{loss}}$ (green). Each set of loaded atoms is prepared and measured 11 times. Panel (c) shows data from a single run, measured in 231~ms of wall-clock time. Panel (d) shows data averaged over 20 runs, measured in 4.6~s of wall-clock time.
    (e) Remaining atom fraction after multiple rounds of initialization and measurement without reloading. 2.5\% of the atoms are lost per reuse.
    }
    \label{fig:fig3}
\end{figure}

\section{Metastable qubit initialization and loss-resolving spin readout}

To use the loaded atoms as qubits, several additional steps are required: light-assisted collisions (LAC) to produce singly-occupied tweezers~\cite{schlosser2001sub}, non-destructive imaging, cooling, and initialization into the $^3P_0$ metastable state (Fig.~\ref{fig:fig3}a). Without substantial optimization, we can perform LAC in \SI{6}{ms}, imaging in \SI{4}{ms} and cooling to $\SI{10}{\mu K}$ in \SI{6}{ms} (App.~\ref{sec:lightassandimg}). To implement high-fidelity imaging, we reduce the size of the tweezer array from 16$\times$16 to 5$\times$5 to suppress background fluorescence induced by the tweezer light in the objective lens~\cite{Saskin2019}. Finally, atoms are optically pumped into  $\ket{1}\equiv\ket{6s6p\,^3P_0,F=1/2,m_F=+1/2}$ in 0.7 ms~\cite{ma2023a} (Fig.~\ref{fig:fig1}c~(i), App.~\ref{sec:experimentsequence}).

To read out the qubit state without blowing atoms out of the trap, we transfer $\ket{0}$ and $\ket{1}$ to $^1S_0$ sequentially and acquire two images. Spin-selective transfer from $^3P_0$ to the ground state is implemented using a two-photon Raman transition to $^3P_2$ followed by optical pumping via $6s6d\,^3D_2$ (Fig.~\ref{fig:fig1}c(i)). The $^3D_2$ state is chosen for its fast decay rate ($\tau = 24$\,ns) and low branching ratio to $^3P_0$ (estimated to be 0.14\%). A similar scheme was demonstrated in Ref.~\cite{lis2023a} using coherent transfer from $^1S_0$ to $^3P_0$ with a clock laser; the optical pumping approach demonstrated here avoids the need for state-insensitive tweezers as the Raman transitions and pumping steps can be performed while blinking off the tweezers~\cite{zhang2025}.

By analyzing both images, there are three possible measurement outcomes: $\ket{0}$ (first image bright), $\ket{1}$ (first image dark and second image bright) and loss (both images dark). The performance is characterized in Fig.~\ref{fig:fig3}b. When preparing the atom in $\ket{0}$, the probability to record the outcome $\{\ket{0},\ket{1},\mathrm{loss}\}$ is $\{0.9759(11),0.0057(5),0.0185(9)\}$. When preparing the atom in $\ket{1}$, it is $\{0.0084(6),0.9529(15),0.0386(13)\}$. The probability to record the correct outcome is $0.9644(13)$, which rises to $0.9927(10)$ when not including loss (App.~\ref{sec:experimentsequence}). 

Using this sequence, we can perform continuous qubit preparation and spin readout of freshly loaded atom arrays at a rate of 30 cycles per second. By imaging each array repeatedly before loading a new array and skipping light-assisted collisions, the measurement rate can be increased to 50 cycles per second. Fig.~\ref{fig:fig3}c shows a loss-resolved Rabi oscillation acquired with 25 sites in only 231 ms, using a single set of loaded atoms. This highlights the power of this approach for fast measurements and calibration in future quantum processors.

\begin{figure}[h!]
    \centering
    \includegraphics[width=0.5\columnwidth]{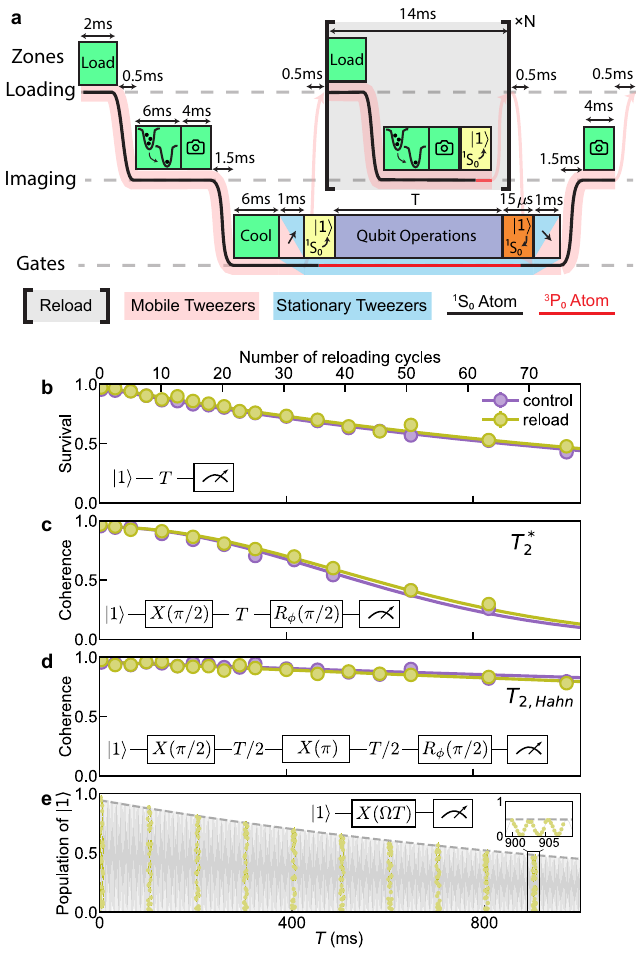}
    \caption{\label{fig:fig4}
    \textbf{Coherence of the $^3P_0$ nuclear spin qubit with continuous reloading.}
(a)~Sequence diagram for coherence measurements with continuous reloading and initialization of fresh $^3P_0$ qubits.
(b)~Measurement of the qubit lifetime in $^3P_0$, with continuous reloading (yellow) and in a control experiment without reloading (purple). In the control experiment, all operations within the gray box in panel (a) are omitted. The fitted $1/e$ decay times are $\tau=\SI{1.30(3)}{s}$ (control) and \SI{1.35(4)}{s} (reloading). 
(c)~Ramsey coherence of individual tweezers, with fitted $1/e$ coherence times of $T_2^*$ = 0.69(2)~s (control) and 0.73(2)~s (reload). The coherence is measured by scanning the phase of the final $\pi/2$ pulse and measuring the fringe visibility, normalized by the atom survival probability.
(d)~Hahn echo measurement, with fitted $1/e$ coherence times of $T_{2,\mathrm{Hahn}}$ = 7(1)~s (control) and 5.4(9)~s (reload).
(e)~Continuous Rabi oscillation while reloading. The dashed line indicates the qubit lifetime from panel (b), showing that the Rabi oscillations are not affected other than loss. 
}
\end{figure}

\section{Coherence preservation and gate operations while reloading}
Finally, we demonstrate that reloading new qubits does not disturb the coherence of existing qubits in $^3P_0$. We introduce a stationary storage array using an SLM~(Fig.~\ref{fig:fig1}c) and transfer a newly loaded batch of qubits to this array.
Immediately after the transfer, the mobile tweezers return to the loading zone and repeat the cycle of loading, parity projection, imaging and qubit initialization. Meanwhile, we perform coherent operations on the $^3P_0$ qubits stored in the stationary tweezers. The stored qubits are transferred back to the mobile tweezers for readout.

In Fig.~\ref{fig:fig4}(b-d), we compare the $^3P_0$ survival and coherence with continuous reloading and in a control experiment where all reloading operations are omitted after loading the initial batch. The results are indistinguishable, demonstrating that the metastable qubit is immune to cooling, light-assisted collision, imaging, and optical pumping light operating on the $^1S_0$ ground state, as anticipated.

The metastable state lifetime in the stationary tweezers is measured to be 1.3 s. This is shorter than the limit from Raman scattering and photoionization (2~s), and vacuum losses (> 15~s). The additional loss arises from technical intensity noise of the transport ODT resulting from phase-to-amplitude conversion by the cavity (App.~\ref{sec:rin}). The $T_2^*$ time for atoms that are not lost is 0.69(2) s, limited by slow magnetic field fluctuations, while the $T_2$ time with a simple Hahn echo is 7(1) s. These measurements place a lower bound on the spin-flip rate of $T_1 > T_2/2 = 3.5$~s, though we expect that $T_1$ is actually much longer because of the absence of Raman scattering for the nuclear spin qubit~\cite{ma2023a,lis2023a}. The long $T_1$ and $T_2$ are compatible with preserving the erasure error bias of the metastable qubit, which is important for reducing the overhead for fault-tolerant quantum computing~\cite{wu2022}.

Finally in Fig.~\ref{fig:fig4}e we show that it is also possible to perform gate operations while reloading. A continuous Rabi drive between $\ket{0}$ and $\ket{1}$ exhibits no decay beyond that imposed by the $^3P_0$ state lifetime.

\section{Discussion and conclusion}

These results upend the usual timescale of optical tweezer experiments, with the 1~ms atom loading time now constituting one of the fastest steps instead of the slowest. As a result, light assisted collisions, non-destructive imaging and cooling now account for over 90\% of the cycle time, which motivates exploring faster implementations of these operations. With realistic improvements~\cite{lis2023a} and by parallelizing these stages in multiple tweezer arrays, it is conceivable to implement several hundred cycles of atom loading, gates, and measurements per second.

These techniques can be extended to create a fully-functional quantum processor by adding rearrangement~\cite{endres2016,barredo2016}, two-qubit gates~\cite{peper2025} and mid-circuit measurement. The measurement protocol demonstrated here can be readily adapted to act on only a subset of the $^3P_0$ qubits by locally addressing the $^3P_0$-$^3P_2$ Raman transfer using local light shifts on the $^3P_2$ state~\cite{burgers2022,Zhang2024Scaled}.

With these additions, it will be possible to explore unlimited-depth circuits with a neutral atom processor. One interesting area of exploration is leakage reduction strategies to preserve logical information across many cycles of atom replacement~\cite{sahay2023a,Chow2024Circuit,perrin2025quantum,baranes2025leveraging}. The fast atom replacement also opens the door to fusion-based strategies explored in photonic quantum computing~\cite{bartolucci2023fusion,sahay2023a}. Examining the subthreshold scaling of repetition codes can probe error floors and correlated errors that may require mitigation~\cite{acharya2023}. Finally, these techniques can reduce the deadtime and increase the data rate in tweezer-based optical atomic clocks~\cite{cao2024multi,shaw2024multi}.

Note: While finalizing this manuscript, another work appeared reporting mid-circuit atom replacement and ancilla qubit reuse using ground-state $^{171}$Yb qubits~\cite{muniz2025}.

\section{Acknowledgments}

We acknowledge helpful conversations with Adam Kaufman, Jon Simon, Lawrence Cheuk, Genyue Liu, Bichen Zhang and Guillaume Bornet, and technical contributions from Mingkun Zhao, Nathan Bonvalet, Martin Poitrinal and Riccardo Pennetta. This work was supported by the Gordon and Betty Moore Foundation (grant DOI 10.37807/gbmf12253), the Army Research Office (W911NF2410358), the National Science Foundation through the CAREER program (PHY-2047620), DARPA MeasQuIT (HR00112490363), the Office of Naval Research (N00014-23-1-2621), and the NSF Center for Robust Quantum Simulation (OMA-2120757).

\section{Competing interests} J.D.T is a co-founder and shareholder of Logiqal, Inc.

\appendix

\section{Experimental apparatus}
\label{sec:experimentappratus}

\begin{figure}[tb]
    \centering
    \includegraphics[width=0.5\columnwidth]{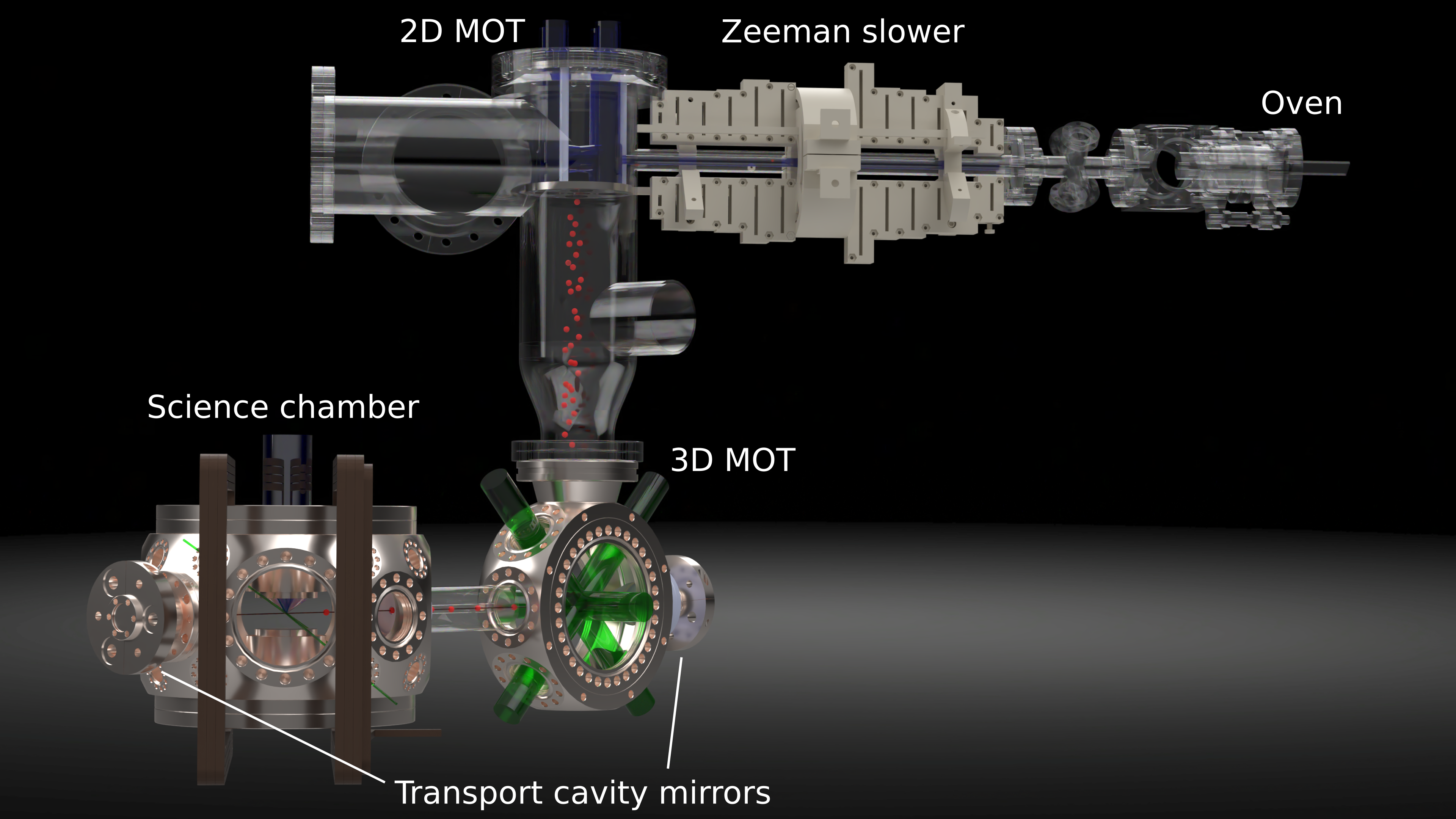}
    \caption{Model of the experiment apparatus.}
    \label{fig:cad}
\end{figure}

\subsection{Continuous Laser Cooling of $^{171}$Yb}

Our experiment begins by supplying cold atoms into the reservoir, where several stages of laser cooling are required, each performed under different magnetic field conditions. As all these stages are operated continuously, no laser or magnetic fields are pulsed or switched in any of these stages.

We create an atomic beam of $^{171}$Yb using an oven from \textsc{AtomSensors}~\cite{Schioppo2012A}. We observed corrosion of the internal nickel gasket in the presence of hot Yb and modified the oven to use a homemade stainless steel gasket, water-jet machined from a piece of annealed 304 stainless-steel sheet. The on-axis flux of $^{171}$Yb from the oven is estimated to be $1.7\times10^{14}$ atoms/s/steradian.

The atomic beam is slowed by a \SI{30}{cm}-long Zeeman slower built from permanent neodymium magnets held in a 3D-printed support. The Zeeman slower operates \SI{-680}{MHz} detuned from the $^1S_0$ to $^1P_1, F =3/2$ transition of $^{171}$Yb. Slowed atoms are loaded into the 2D MOT at a detuning of \SI{-38}{MHz}, with a magnetic field gradient of $\SI{60}{G/cm}$ generated by two permanent neodymium magnets held in a 3D-printed structure outside of the vacuum chamber. Compensation magnets are placed between the 2D MOT and 3D MOT chamber to minimize the crosstalk of the 2D MOT magnetic field to the 3D MOT region.

The vertical distance between the 2D MOT and 3D MOT is \SI{35}{cm}. Atoms free falling from the 2D MOT region would gain a velocity of \SI{2.6}{m/s} arriving at the 3D MOT region, exceeding the capture velocity of a typical single-frequency narrow line MOT operating on the $^1S_0\rightarrow{}^3P_1$ intercombination line of $^{171}$Yb. To address this, we rapidly scan the detuning of the 3D MOT light from \SI{-28}{MHz} to \SI{-0.15}{MHz} using a sawtooth wave with a period of 2.75~$\mu$s.
The 3D MOT capture rate is $2\times10^{6}$ atoms/s.

\begin{figure*}[h!]
    \centering
    \includegraphics[width=\textwidth]{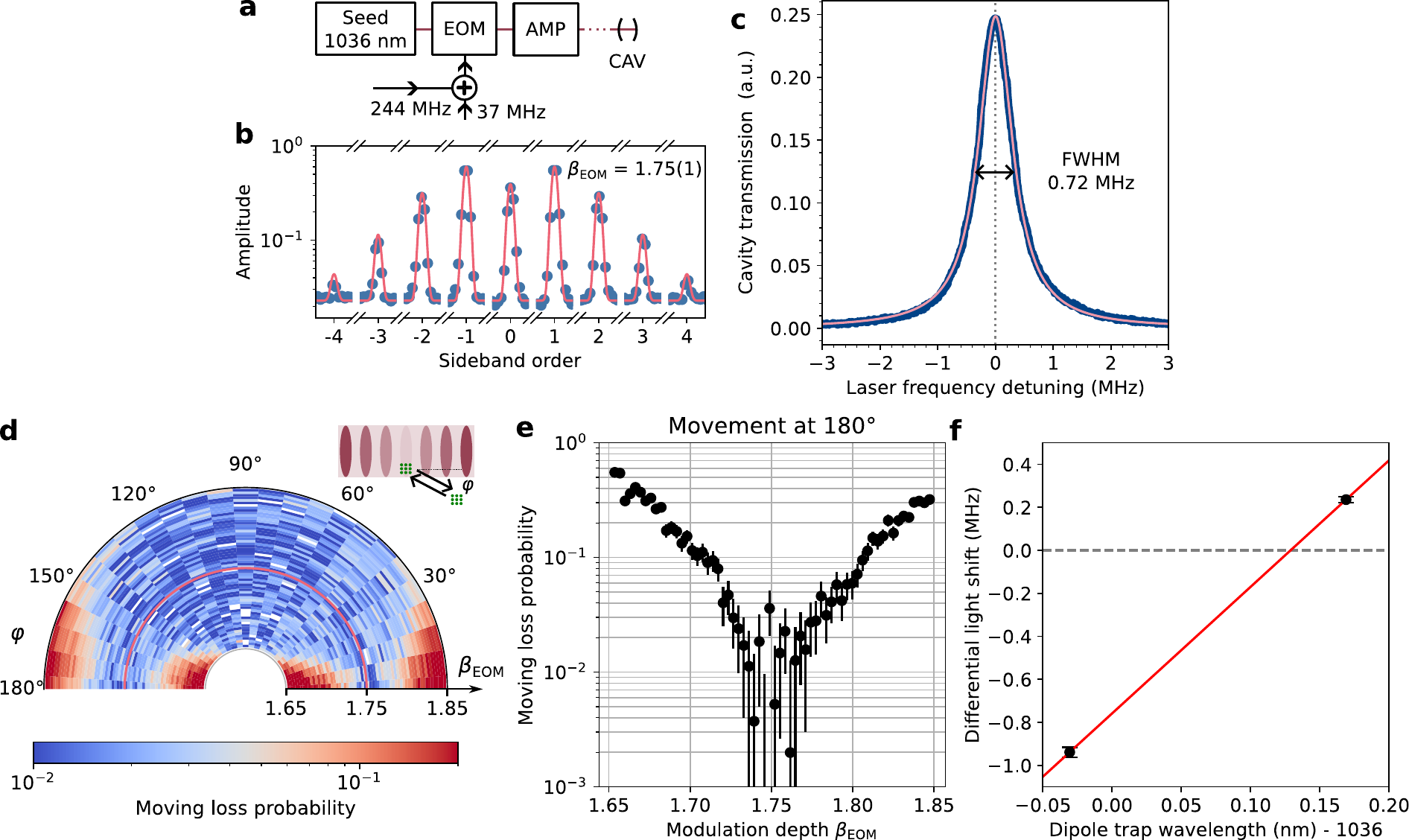}
    \caption{\textbf{Details of the transport cavity operation. }
    (a) Schematic laser setup for the cavity-enhanced dipole trap. A Yb-doped fiber amplifier is seeded by a Yb-fiber laser. The output of the fiber laser is phase modulated using an EOM at two RF frequency for PDH locking (\SI{37}{MHz}) and standing-wave mitigation ($\nu_{\text{FSR}}=\SI{244}{MHz}$).
    (b) Blue points: Relative amplitudes of sidebands of the \SI{244}{MHz} EOM drive, obtained from a beat-note measurement on a photodiode of the cavity transmission with a frequency shifted (\SI{100}{MHz}) portion of the seed laser. Pink line: Analytical expression for the EOM sideband amplitudes at a modulation depth of $\beta_{\mathrm{EOM}} = 1.75(1)$. 
    (c) Blue: Cavity transmission signal with the FSR modulation applied, as a function of the laser frequency.
    Pink: Lorentzian line fit to the cavity transmission signal with a full-width half maximum (FWHM) linewidth of \SI{0.72}{MHz}. 
    (d) Round-trip loss of atoms during movement in the computation zone as a function of the EOM modulation depth $\beta_\mathrm{EOM}$ and angle of movement $\varphi$ relative to the cavity-enhanced dipole trap axis. The pink line indicates the optimal modulation depth of $\beta_{\mathrm{EOM}} = 1.75$.
    (e) Moving round-trip loss probability along cavity-enhanced ODT as a function of the EOM modulation depth $\beta_\mathrm{EOM}$.
    (f) Black points: Differential light shifts on the $\ket{6s^2\,^1S_0,F=1/2,m_F=\pm 1/2}\rightarrow\ket{6s6p\,^3P_1,F=3/2,m_F=\pm 1/2}$ transition as a function of the dipole trap wavelength. 
    Red line: Interpolation between the two measured differential light shifts. 
}
    \label{fig:cavitymod}
\end{figure*}

\subsection{Transport Cavity}
\label{sec:tranportcav}

Efficient loading from the 3D MOT into the transport ODT and subsequent loading from the ODT into the optical tweezers are facilitated by choosing an ODT wavelength that is magic for the intercombination line~\cite{Cooper2018Alkaline,Norcia2018Microscopic,Saskin2019}. We select \SI{1036}{nm}~\cite{Zheng2020Magic} for this purpose. This wavelength is also convenient because of the availability of high-power lasers, and because only small frequency adjustments are needed to reach the magic condition for either $^{174}$Yb or $^{171}$Yb. However, at \SI{1036}{nm} the dipole polarizability is relatively small at approximately $160\,a_0^3$ \cite{Zheng2020Magic}. To achieve a trap depth over $\SI{100}{\mu K}$ against gravity across the entire \SI{35}{cm} transport distance, a laser power of \SI{1.1}{kW} would be required in a single-beam trap. To avoid excessively high laser powers, we use a moderate finesse in-vacuum build-up cavity in Fabry-P\'erot configuration.

The power enhancement factor $\mathcal{E}$ of a Fabry-P\'erot cavity (up to losses) is given by
\begin{equation}\label{eq:powerenhancement}
    \mathcal{E} = P_\mathrm{circ}/P_\mathrm{in} = \mathcal{T}_1 / (1-\sqrt{\mathcal{R}_1\mathcal{R}_2})^2 \approx \frac{2\mathcal{F}}{\uppi}\frac{\mathcal{T}_1}{\mathcal{T}_1+\mathcal{T}_2}
\end{equation}
where $P_\mathrm{circ}$ is the circulating power in the cavity, $P_\mathrm{in}$ is the incident power on the incoupling mirror M1 and $\mathcal{T}_i$ ($\mathcal{R}_i$) is the transmission (reflection) of mirror $i$. The last equality holds approximately when $\mathcal{T}_i\ll1$. The enhancement factor for strongly over-coupled cavity ($\mathcal{T}_1\gg \mathcal{T}_2$) is twice that of a critically coupled ($\mathcal{T}_1=\mathcal{T}_2$) cavity. We therefore choose $\mathcal{R}_1=0.98$ and $\mathcal{R}_2=0.9999$ yielding power-enhancement factor of $\mathcal{E}\approx196$ and cavity finesse of $\mathcal{F}=309$. 

The mirrors have a radius of curvature of $R_{oc}=\SI{500}{mm}$, producing a close to optimal Gaussian beam waist of $\SI{280}{\mu m}$ at the center of the cavity that balances intensity and diffraction. The mirror separation is \SI{61}{cm} yielding a free spectral range (FSR) of \SI{244}{MHz}. For alignment purposes, the cavity mirrors are mounted on small sections of flexible bellows, allowing for control of tilt and translation by threaded rods and nuts outside the vacuum chamber. The transport cavity axis is oriented at \SI{45}{\degree} to the tweezer bias magnetic field coils (see Fig.~\ref{fig:cad}) to allow for unrestricted optical access for gate and optical pumping beams both perpendicular and parallel to the quantization axis.

The cavity is driven by an amplified Yb fiber laser (seed: \textsc{NKT}, BASIK Y10, amplifier: \textsc{IPG, YAR-20-1030-LP-SF}), delivering \SI{18}{W} to the cavity. From the transmitted power, we estimate the intra-cavity power to be approx. \SI{3.3}{kW}. The laser frequency is locked to the fixed-length cavity using the Pound-Drever-Hall (PDH) technique with sidebands applied by a fiber electro-optic modulator (EOM) ($f_\mathrm{PDH} = \SI{37}{MHz}$, \textsc{exail}, NIR-MPX-LN-0.1, Fig.~\ref{fig:cavitymod}a).

The cavity standing wave hampers atom transport, but we can suppress it significantly by driving multiple longitudinal modes of the cavity spaced by the FSR. The FSR modulation is applied using the same EOM that generates the PDH sideband (Fig.~\ref{fig:cavitymod}a). It is possible to cancel the standing wave exactly at two symmetric points lying at the tweezer region and close to the 3D MOT (Fig.~\ref{fig:fig1}b). A representative cavity transmission signal as a function of the laser frequency is presented in Fig.~\ref{fig:cavitymod}c. By tuning the EOM modulation frequency to match the cavity's FSR, all sidebands become resonant with the cavity at the same time when the carrier is locked to a cavity resonance. This allows us to precisely determine the FSR by maximizing the cavity transmission signal.

The position along the cavity axis where the standing wave is exactly nulled varies with phase modulation depth $\beta_\mathrm{EOM}$. To precisely find $\beta_\mathrm{EOM}$ matching the tweezer loading location, we probe the atom transport loss as a function of the moving angle and $\beta_\mathrm{EOM}$ (Fig.~\ref{fig:cavitymod}d-e). We find a clear minimum in the moving loss probability when moving along the cavity axis at $\beta_\mathrm{EOM} = 1.75(1)$, and operate the cavity at this point. The $\beta_\mathrm{EOM}$ is calibrated by fitting Bessel functions to the measured intensity distribution of the laser sidebands obtained by recording the beat signal of the cavity transmission with a frequency-shifted (\SI{100}{MHz}) portion of the seed laser. A typical measurement for the optimal $\beta_\mathrm{EOM} = 1.75(1)$ is presented in Fig.~\ref{fig:cavitymod}b.

Furthermore, the restoring force of the 3D MOT is larger than the potential gradient of the transport ODT along its axial direction, prohibiting the atoms from leaving the MOT region and flowing towards the science chamber. We partially mask the MOT beams that are in the plane of Fig.~\ref{fig:fig1}a and intersecting the transport ODT, allowing the optical gradient force of the ODT to pull atoms towards the science chamber.

Based on the literature, we expect the magic wavelength for the $^{171}$Yb $\ket{6s^2\,^1S_0,F=1/2,m_F=\pm1/2}\rightarrow\ket{6s6p\,^3P_1,F=3/2,m_F=\pm1/2}$ transition to lie at \SI{1035.99(3)}{nm} (adapting the reported magic wavelengths for $^{174}$Yb \cite{Zheng2020Magic} to $^{171}$Yb following Ref.~\cite{ma2022}). By measuring the light shifts in the absorption spectrum of the molasses, we find that the light shift vanishes when the laser is set to \SI{1036.13(1)}{nm} (Fig.~\ref{fig:cavitymod}f). However, the magic wavelength lies very close to the $6s6p\,^3P_1\rightarrow5d6s\,^1D_2$ transition at \SI{1032.45}{nm} \cite{NIST_ASD}. We have found that the polarizability is very susceptible to the ASE background from the Yb fiber amplifier, and the operational magic wavelength depends sensitively on how the output light is filtered. We currently use a 3 nm bandwidth angle-tuned dichroic filter, but we are not confident that this removes all of the ASE.

To prevent atom flux from accumulating on the cavity mirrors, we blow away atoms just before each mirror with a resonant 399 nm laser. We have observed no change in the finesse over 1 year of daily operation. It is possible that the residual cavity standing wave itself is sufficient to prevent atoms from reaching the mirrors, but we have not tested this hypothesis.

\subsection{Molasses and Tweezer Array}
In the science chamber, a 1D optical molasses operating on the intercombination line with cooling beams applied at a 25.6$^\circ$ angle to the ODT captures atoms into a $\SI{500}{\mu m}$ long, $\SI{100}{\mu m}$ wide region under the objective field of view (Fig.~\ref{fig:fig1}c~(ii)). The magnetic bias field in the science chamber (8.5~Gauss) splits the $\ket{^3P_1,\,m_F}$ levels by $12\,\mathrm{MHz}$~(66$\Gamma_{^3P_1}$). Therefore, we apply a molasses with two frequencies to address the $\ket{^3P_1,\,m_F=\pm1/2}$ states separately. When operated continuously, the molasses loading rate is $1.6\times 10^{6}$~atoms/s, and we reach a steady-state atom number of $4.3\times 10^6$ atoms with a peak density of 8.0$\times 10^{11}$~$\mathrm{atoms/cm^3}$ and a radial temperature of 87(4)~$\mu$K. The steady-state atom number is limited by light-assisted two-body loss with an estimated coefficient of $\beta = 6.1(3)\times 10^{-13}$~$\mathrm{cm^3/s}$ (App.~\ref{sec:loadingdynamics}1).

The mobile tweezer array is a 2D array created by a pair of crossed AODs (\textsc{AA Optoelectronic}, DTSX-400-488). To accommodate the large distance ($\sim\SI{280}{\mu m}$) required for moving the mobile tweezer array between loading zone and computation zone, we perform the translation using a dual-axis galvo mirror scanner (\textsc{Thorlabs}, QS10XY-AG). The galvo warms up considerably when operating at high duty cycle, which changes its alignment. Stable long-term performance is maintained by tracking the array position on a camera and adjusting the programmed trajectory during the experiment. The stationary tweezer array is created by a liquid crystal on silicon spatial light modulator (LCoS-SLM, \textsc{Santec}, SLM-300).

The wavelength of the tweezer laser is $\sim\SI{488}{nm}$, which is a magic wavelength of the $^1S_0\rightarrow\ket{{}^3P_1,F=3/2,|m_F|=1/2}$ transition in $^{171}$Yb~\cite{ma2022}. The laser beams of the two tweezer arrays are combined using a polarization beam splitter cube, in which the polarization of the mobile tweezer array is aligned with the magnetic field to enable magic trapping for tweezer loading, high fidelity imaging and cooling. The combined tweezer light is focused into the tweezer chamber through a custom microscope objective lens (\textsc{Special Optics}) with a numerical aperture of 0.55. The science chamber is a metal chamber with re-entrant viewports. The microscope objective is interferometrically aligned to the re-entrant vacuum window, referenced using an in-vacuum reflective bearing ball. 
We use an optical tweezer power of 2.4 mW/trap for loading and transport large arrays, and 4.5 mW/trap for high-fidelity imaging, measured at the input to the objective.
The \SI{556}{nm} photons scattered from the atoms during imaging are collected by the same microscope objective lens, separated using a dichroic mirror and focused onto a low noise CMOS camera (\textsc{Hamamatsu}, ORCA C15550-20UP).

\section{Loading dynamics}
\label{sec:loadingdynamics}

\subsection{Molasses loading}
\label{subsec:molassesloading}

The optical molasses used to load the reservoir is generated by two laser tones spaced by $\SI{12}{MHz}$ to match the $^1S_0\rightarrow\ket{^3P_1,m_F=-1/2}$ and $^1S_0\rightarrow\ket{^3P_1,m_F=+1/2}$ transitions in the $\SI{8.5}{G}$ bias magnetic field. Each tone is detuned $4.4\Gamma_{\text{556}}$ red detuned from resonance with an intensity of $4 I_{\text{sat,mol}}$. Note that the linearly polarized molasses beam can drive any of the $\sigma^+$, $\sigma^-$ and $\pi$ transitions, because of the $\SI{62}{\degree}$ angle between its k-vector and the bias magnetic field, and thus $I_{\text{sat,mol}} = 0.42$~mW/cm$^2$ is the saturation intensity accounting for the polarization of the laser beam. $\Gamma_{\text{556}}=2\pi\times\SI{182.4}{kHz}$ is the nature linewidth of the $^1S_0\rightarrow {}^3P_1$ transition. 

The loading dynamics of the molasses is described by the equation
\begin{equation}
    dn(t)/dt = -\Gamma_{\text{atom}} n(t) - D\beta n(t)^2 + D\phi_0/V_{\mathrm{eff}},
    \label{eq:molasses_loading_eq}
\end{equation}
where $\Gamma_{\text{atom}}$ is the single atom loss rate, $\beta$ is the light-assisted two-body loss coefficient, $D$ is the molasses duty cycle, $\phi_0$ is the incoming atom flux, and $V_{\mathrm{eff}}$ is the effective volume of the molasses. 
We fit the molasses loading curves in Fig.~\ref{fig:fig2}b with this equation and extract the parameters $\Gamma_{\text{atom}} = 0.156(8) \,\mathrm{s^{-1}}$, $\beta = 6.1(3)\times 10^{-13}$~$\mathrm{cm^3}$/s, $\phi_0 = 1.60(4)\times 10^{6}$~atoms/s, and $V_{\mathrm{eff}} = 6.6(4)\times 10^{-6}$~$\mathrm{cm^3}$.
The steady state molasses density can be calculated using this equation as
\begin{equation}
    n_{\infty} = \frac{-\Gamma_{\text{atom}} + \sqrt{\Gamma_{\text{atom}}^2 + 4\beta \phi_0 D^2/V_{\mathrm{eff}}}}{2\beta D},
    \label{eq:molasses_nss}
\end{equation}
and the calculated $n_{\infty}$ is plotted as the black dashed line in Fig.~\ref{fig:fig2}c.

\begin{figure*}
    \centering
    \includegraphics[width=0.6\textwidth]{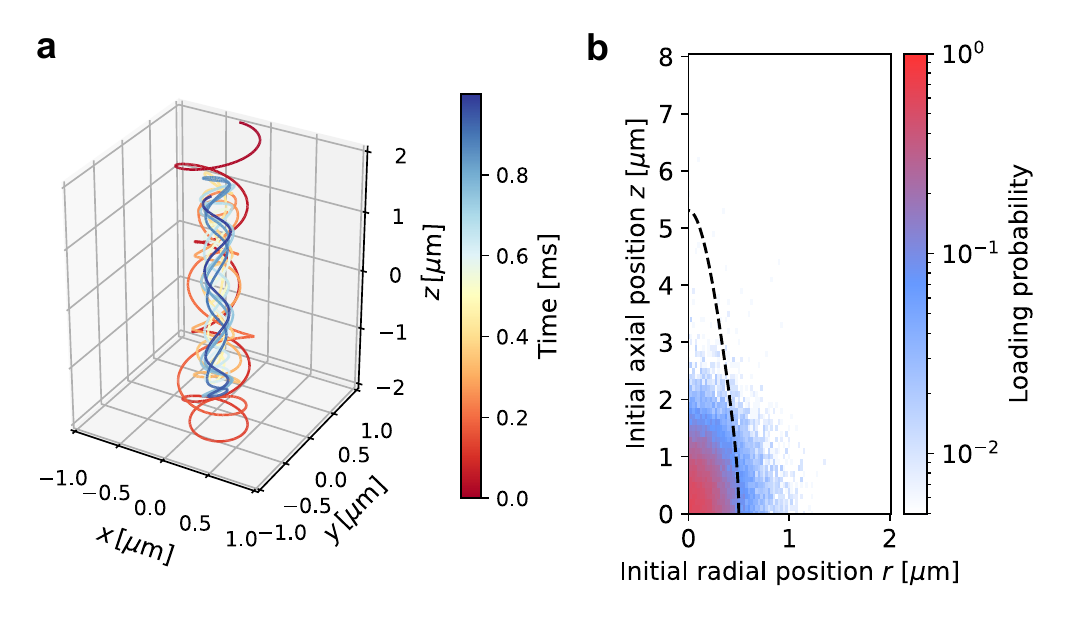}
    \caption{\textbf{Tweezer loading simulation.}
    (a) Representative trajectory of an atom that is captured and remains bound by the trap.
    (b) \SI{1}{ms} loading probability versus initial axial (\(z\)) and radial (\(r\)) position relative to tweezer focus. Only atoms initialized within a distance of one waist from tweezer focus reach appreciable loading probability. The dashed line indicates $1/e^2$ intensity contour of the tweezer.
    }
    \label{fig:loadingsim}
\end{figure*}

\subsection{Tweezer Loading Simulation}
\label{sec:loadingsimulation}
To calculate the tweezer loading rates from the optical molasses, we run a Monte-Carlo simulation of the classical trajectory of the atoms in 3D. The force on the atom includes the position dependent optical dipole trapping force provided by the transport ODT and the optical tweezers. To simulate the Doppler cooling effect, we stochastically apply momentum kicks from photon scattering. The momentum kick probability is obtained from the steady state solution of the optical Bloch equation that includes four levels $\ket{^1S_0,m_F=\pm1/2}$ and $\ket{^3P_1,m_F=\pm1/2}$, using molasses beam parameters listed in the previous section. The simulation is divided into two stages. First, we simulate the molasses loading to get the steady-state velocity distribution for atoms in the molasses. Then, we use this distribution to simulate the dynamics of loading individual tweezers.

The molasses simulation predicts unequal temperatures in the radial and axial direction of $\SI{77}{\mu K}$ and $\SI{33}{\mu K}$, respectively, because the molasses beam has a much larger projection onto the axial direction. The radial temperature matches the experimental observation, but we do not have a way to directly probe the axial temperature in the experiment.

We simulate the tweezer loading by initializing atoms at different positions relative to the tweezer focus, with the velocity drawn from a Maxwell-Boltzmann distribution defined by the temperature derived in the first stage simulation and a real-space density corresponding to the measured reservoir density. For each initial position, 500 trajectories are simulated to calculate the tweezer loading probability after a certain time of evolution.
An atom is determined to be loaded if the sum of its kinetic energy and potential energy below a somewhat arbitrarily chosen energy threshold $E_{\text{tot}} < E_{\text{threshold}} = 0.5 E_{\text{trap\ depth}}$. At a reservoir atom density of $n = 4\times10^{11}\mathrm{/cm^3}$ (corresponding to 50\% duty cycle molasses operation in Fig.~\ref{fig:fig2}c), the predicted loading probability is 0.45 at 1 ms loading time, which closely matches the experimental value.

In Fig.\ref{fig:loadingsim}b, we show the loading probability at $t=1$ ms, as a function of the atom position at $t=0$. Only atoms that start within one beam waist of the tweezer have any significant loading probability. While the characteristic velocity in the molasses is several cm/s and typical atoms travel tens of microns in 1 ms, atoms that start far away pass through the tweezer before they can be cooled. This suggests that it should be possible to load very dense arrays without affecting loading rate.

\section{Parity projection, non-destructive imaging and cooling}
\label{sec:lightassandimg}

\begin{figure*}
    \centering
    \includegraphics[width=\textwidth]{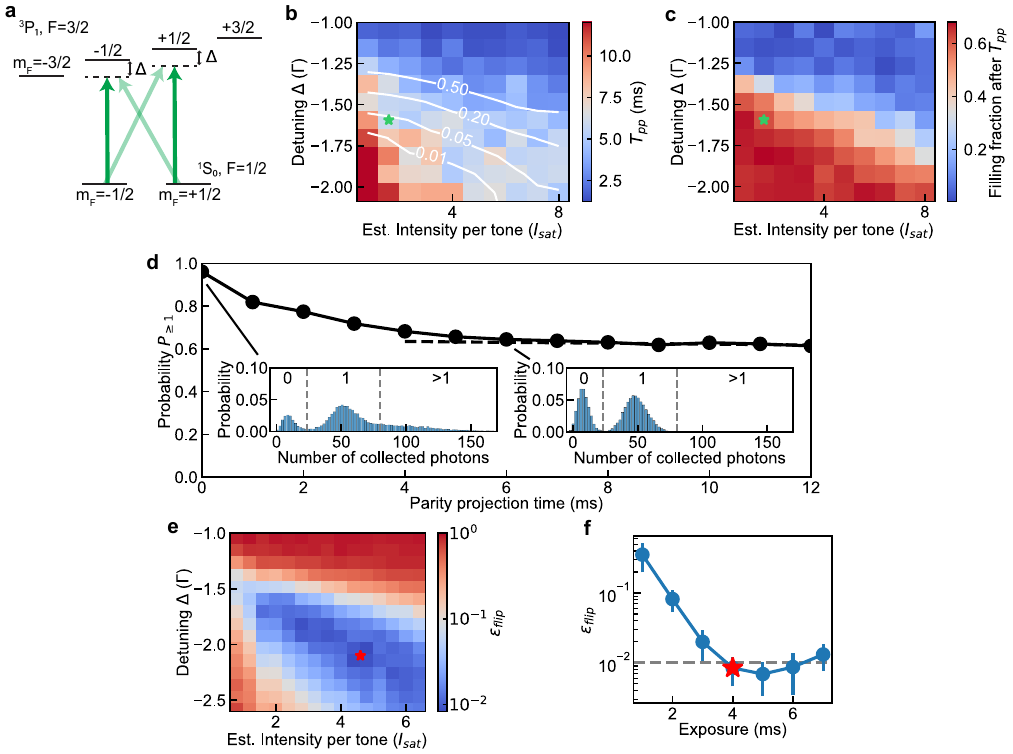}
    \caption{
    \textbf{Parity projection and imaging.} 
    (a) Relevant energy levels during parity projection and imaging.
    (b) Target parity projection time $T_{pp} = 3/\Gamma_{>1}$, where $\Gamma_{>1}$ is the loss rate of more than one atoms in a tweezer. $T_{pp}$ is plotted as a function of laser detuning $\Delta$ from the $\ket{6s6p\,^3P_1,\,m_F=\pm 1/2}$ levels and the estimated intensity per frequency tone of the laser beam.  
    White contours indicate the single-atom loss percentage over the duration of $T_{pp}$. 
    (c) Single-tweezer filling fraction after parity projection of duration $T_{pp}$.
    The green star in panels (b) and (c) denotes the parameters used in daily operation and in panel (d).
    (d) Probability of at least one atom loaded in a single tweezer $P_{\geq 1}$ as a function of parity projection time. 
    Insets show histograms from images taken after 0 ms and 6 ms of parity projection, respectively.
    (e) Normalized probability of result flips in two consecutive images $\epsilon_{\text{flip}} = p_{bd}/p_b + p_{db}/p_d$ as a function of laser detuning $\Delta$ and the estimated intensity per frequency tone of the laser beam. 
    (f) $\epsilon_{\text{flip}}$ as a function of imaging exposure time. 
    The red star in panels (e) and (f) indicates the parameters used in daily operation and in panel (d).
    All data in panels (b-f) were taken at a tweezer trap depth of 3.6 MHz.
    }
    \label{fig:figImgPP}
\end{figure*}

We perform parity projection to achieve a sub-Poissonian distribution in our tweezers using a light-assisted collision~(LAC) pulse after loading. 
The LAC pulse is applied using a local laser beam (local beam) that only illuminates the imaging and storage zone, but not the loading zone where the molasses is. 
We explore the parameter space of light-assisted collision in search for a condition that provides a fast light-assisted collision rate for multiple atoms, while maintaining a high survival rate of single atoms. To characterize the loss probabilities, We apply a LAC pulse of variable length between two images and extract the decay rate of the number of events where photon counts exceed a threshold where more than one atom is loaded~($\Gamma_{>1}$). The imaging pulses are \SI{2}{ms}, which is shorter than a normal imaging pulse (\SI{4}{ms}), to keep the LAC effect small during the first image. To probe the single atom loss rate, we apply a known working LAC pulse before the first image, and extract the decay rate of the survival fraction of single atoms~($\Gamma_{1}$).

We set the target of parity projection to reducing the probability of having multiple atoms to $e^{-3}$ of that after loading, corresponding to parity projection time $T_{pp}=3/\Gamma_{>1}$. It is possible to achieve faster parity projection by choosing a higher intensity or closer detuning (Fig.~\ref{fig:figImgPP}b), in exchange of higher single-body loss probability~(Fig.~\ref{fig:figImgPP}b, white contours) thus lower filling (Fig.~\ref{fig:figImgPP}c). We operate our parity projection at the parameters indicated by the green star. The parity projection takes \SI{6}{ms}, with the single-body loss probability of $5\%$, and a saturated filling fraction of $60\%$ (Fig.~\ref{fig:figImgPP}d).

We also explore the parameter space for optimal imaging~(Fig.~\ref{fig:figImgPP}e). The local beam is used for tweezer imaging. For each tweezer, we analyze a 3-by-3 region of interest and determine the tweezer occupancy based on a photon count threshold. Images with counts above the threshold are classified as bright ($^1S_0$ atom present), and those below as dark (no $^1S_0$ atom).
By taking two consecutive images and calculating the normalized probability of the results being different $\epsilon_{\text{flip}} = p_{bd}/p_b + p_{db}/p_d$, we optimize our imaging taking into account both identification error and loss. 
Our daily operation uses an imaging time of \SI{4}{ms}~(Fig.\ref{fig:figImgPP}f). With $\epsilon_{\text{flip}}$ of 0.8(4)~\%, we reach a good balance of short imaging time, high imaging fidelity and low atom loss.

After imaging, a \SI{6}{ms} long cooling pulse is applied to reduce the atom temperature to approximately $\SI{10}{\mu K}$ \cite{blodgett2023imaging}. The cooling pulse requires both local beam and molasses beam, where each beam contains the same frequency that is \SI{2.0}{MHz} blue detuned from the $\ket{{}^1S_0,\,m_{\text{F}}=+1/2}\rightarrow \ket{{}^3P_1,\, m_{\text{F}}=+1/2}$ transition. The local beam has an intensity of $30 I_{\text{sat,556}}$ and the molasses beam has a intensity of $90 I_{\text{sat,556}}$.

\section{Metastable qubit initialization and measurement}

After parity projection, taking an identification image, and cooling, the atoms are ready to be prepared to the $^3P_0$ qubit subspace using optical pumping. 
To avoid light shifts and loss induced by anti-trapping of $^3P_2$ states during the optical pumping and depumping process, it is beneficial to turn off the optical tweezers during these operations. Therefore, we switch to operating the tweezers in a trap intensity-modulated mode~\cite{hutzler2017eliminating,zhang2025} with a pulse width modulated using a $50\%$ duty cycle square wave with a period of $\SI{1.8}{\mu s}$, providing $\SI{900}{ns}$ trap off time per cycle for operations. 
The trap intensity modulation can be used in both the mobile tweezers and the stationary tweezers, enabling pumping and depumping operations in both arrays.
To ensure low atom loss when switching on trap intensity modulation in the mobile tweezers, the average tweezer intensity is kept constant while the intensity modulation duty cycle is gradually ramped from 100\% to 50\%.
To ensure low loss transfer between the mobile tweezer array and the stationary tweezer array, the mobile tweezer array intensity is ramped down while ramping up the average intensity of the intensity-modulated stationary tweezers. 

\begin{figure*}
    \centering
    \includegraphics[width=0.8\textwidth]{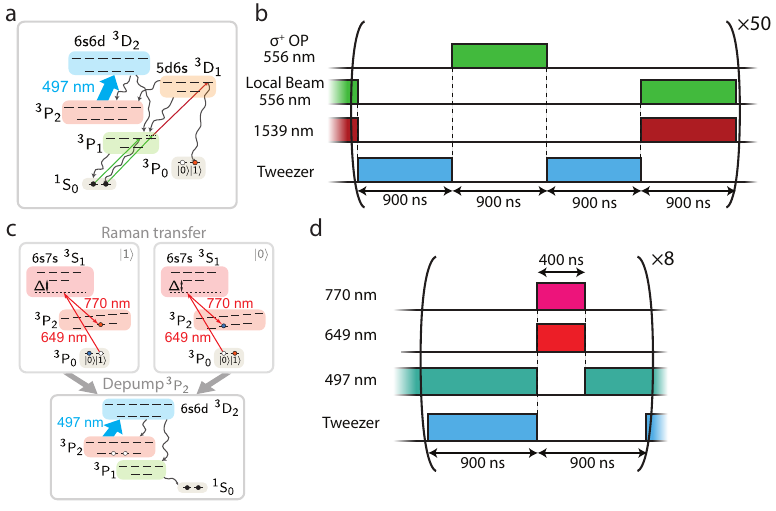}
    \caption{(a) Relevant energy level diagram and (b) timing diagram for optical pumping to $\ket{1}$. (c) Relevant energy level diagram and (d) timing diagram for qubit readout.
    }
    \label{fig:readout_schematic}
\end{figure*}

To prepare the atoms in the $\ket{^3P_0,m_F=+1/2}$ qubit state ($\ket{1}$), we employ two stages of optical pumping. In the first stage, we apply a global $\sigma^{+}$ optical pumping beam with a well-defined circular polarization that drives the $\ket{^1S_0,m_F=-1/2}\rightarrow\ket{^3P_1,m_F=+1/2}$ transition to prepare the atoms in $\ket{^1S_0,m_F=1/2}$ state. In the second stage, we coherently excite atoms to $5d6s~^3D_1$ using a two-photon Raman transition $\ket{^1S_0,m_F=1/2}\rightarrow\ket{^3P_1,m_F=3/2}$ (\SI{556}{nm}) and $\ket{^3P_1,m_F=3/2}\rightarrow \ket{5d6s~^3D_1,m_F=3/2}$ (\SI{1539}{nm}), with a single-photon detuning of \SI{10}{MHz} (Fig.~\ref{fig:readout_schematic}a). The single-photon Rabi frequency is \SI{3.6}{MHz} for \SI{556}{nm} and \SI{2.8}{MHz} for \SI{1539}{nm}. 
The $5d6s~^3D_1$ state decays into both $6s6p~^3P_0$, $6s6p~^3P_1$, and $6s6p~^3P_2$. $6s6p~^3P_1$ decays back to $^1S_0$ and is pumped again, and $6s6p~^3P_2$ is constantly depumped through the $6s6d~^3D_2$ state.
The \SI{556}{nm} transition is driven by a local beam that illuminates only the tweezer region but not the molasses region. 
Together with a global \SI{1539}{nm} beam, we selectively pump atoms in the tweezers into $^3P_0$ state without affecting the atoms in the molasses. Atoms in the molasses are further protected by the ODT light shift on the $5d6s~^3D_1$. 
These two stages of optical pumping happen during the trap off period to avoid light shift from the tweezers. The pumping process consists of 50 cycles of each stage, as shown in Fig.~\ref{fig:readout_schematic}b.

The tensor light shift induced by the \SI{1036}{nm} ODT mixes the $\ket{m_F=-1/2}$ and $\ket{m_F=+3/2}$ Zeeman sublevels of the $5d6s~^3D_1$ state, allowing some population to decay into the $\ket{^3P_0, m_F=-1/2} (\ket{0})$ state following optical pumping. In this work, we opt to selectively depump $\ket{^3P_0,m_F=-1/2}$ state using the method described below, followed by another round of optical pumping. This process is repeated twice to ensure high spin purity in the $^3P_0$ subspace. 
Note that decay into $|0\rangle$ can be supressed either by applying larger magnetic field, or by turning off the ODT for tens of microseconds during optical pumping without losing the molasses atoms.
With future upgrades, we expect a single round of optical pumping to suffice for achieving high spin purity.

To spin-selectively depump the atoms from $\ket{^3P_0,m_F=\pm1/2}$ qubit states, we drive a coherent two-photon Raman transition using $\ket{^3P_0,m_F=\pm1/2}\rightarrow\ket{6s7s~^3S_1,m_F=\pm1/2}$ (\SI{649}{nm}) and $\ket{6s7s~^3S_1,m_{F}=\pm1/2}\rightarrow\ket{^3P_2,F=3/2,m_{F}=\pm1/2}$ (\SI{770}{nm}), where both beams drive a $\pi$ transition (Fig.~\ref{fig:readout_schematic}c). The detuning from the intermediate state is $\sim\SI{12}{GHz}$. We obtain a two-photon Rabi frequency of \SI{1.08}{MHz}. Due to the large $g$-factor of the $\ket{^3P_2,F=3/2}$ state ($g=1.8$), the two-photon resonance for the two spin states are different by \SI{22}{MHz} at an \SI{8.5}{G} magnetic field, which allows the $\ket{0}$ and $\ket{1}$ transitions to be spectrally separated.

Finally, the atoms in $^3P_2$ are depumped back to the ground state using the $\ket{^3P_2, F=3/2}\rightarrow\ket{6s6d~^3D_2,F=3/2}$ and $\ket{^3P_2, F=5/2}\rightarrow\ket{6s6d~^3D_2,F=5/2}$ transitions. These two transitions are spaced by $\sim\SI{136}{MHz}$, which can be easily addressed using acousto-optic modulators. 
Since $^3P_2$ state is anti-trapped at our tweezer wavelength, the Raman transfer and depumping steps are also carried out during the tweezer trap off period~(Fig.~\ref{fig:readout_schematic}d). 

Compared to a previous implementation of depumping through the $6s7s~^3S_1$ state~\cite{ma2023a}, the $6s6d~^3D_2$ state cannot decay directly to $6s6p\,^3P_0$, enabling spin readout without blowing out one of the spin states. However, the $6s6d~^3D_2$ state can decay to $^3P_0$ indirectly through a multi-step decay involving the $6s7p$ states. To estimate the branching ratio, we calculate the matrix elements using a single-active electron approximation, finding the reduced matrix elements $\langle6s6d||r||6s6p\rangle = 2.1\,a_0$, and $\langle6s6d||r||6s7p\rangle = 12.5\,a_0$. This results in a predicted branching ratio from $6s6d\,^3D_2$ to $6s6p\,^3P_0$ of 0.14\%, which results in a 0.19\% probability of repopulating $6s6p\,^3P_0$ while depumping $6s6p\,^3P_2$.

\label{sec:experimentsequence}

After preparing $\ket{0}$ and $\ket{1}$, the probability to record the correct qubit state in the three-outcome measurement is $P_{(0|0)} = 0.9759(11)$ and $P_{(1|1)} = 0.9529(15)$. When corrected for atom loss, the measurement fidelity is $P_{(0|0)}/(1-P_{\text{loss}}) = 99.43(14)\%$ and $P_{(1|1)}/(1-P_{\text{loss}}) = 99.11(20)\%$.

The probability to observe no atom after preparing $\ket{0}$ is attributed to loss during or after the image to determine the initial occupancy (0.7\%) and loss during the qubit initialization and depump cycle (1.0\%). The latter quantity has been improved to 0.6\% in recent work~\cite{zhang2025}.

The additional loss when preparing $\ket{1}$ results from Raman scattering and photoionization during the first image (2\%, the $^3P_0$ lifetime in the imaging trap depth is \SI{350}{ms}). This can be improved by using a different trapping wavelength or a shorter exposure time.

The remaining atom discrimination errors are attributed to $^3P_0$ decay to $^1S_0$ during the first image (0.35\%), and off-resonant scattering from the two-photon Raman drive (measured to be $9\times 10^{-5}$/pulse, or 0.07\% error during a readout). Additionally, the $\ket{0}$ readout pulses can drive Rabi oscillation from $\ket{1}$ to $^3P_2$ due to off-resonant excitation or polarization impurity, with a measured probability of $3.6\times 10^{-4}$/pulse (0.29\% total probability), and the $^3D_2$ state can decay back to $^3P_0$ with a low probability (est. 0.14\%). The dominant errors in this list can be improved with better control of the Raman drive.

\begin{figure}[tb]
    \centering
    \includegraphics[width=0.5\textwidth]{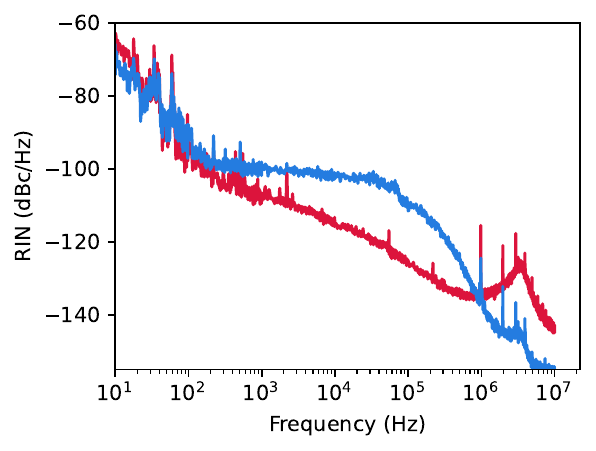}
    \caption{Relative intensity noise (RIN) of the \SI{1036}{nm} ODT laser, measured before (red) and after (blue) the optical cavity.}
    \label{fig:RIN1036}
\end{figure}

\section{Relative intensity noise of transport cavity}
\label{sec:rin}

We attribute the anomalously short atom lifetime in the tweezers to mechanical heating from intensity noise in the transport ODT. In Fig.~\ref{fig:RIN1036} we show measurements of the relative intensity noise (RIN) on the \SI{1036}{nm} after the amplifier and on the transmission through the cavity when the laser is locked to the cavity. In the frequency range between \SI{1}{kHz} and \SI{1}{MHz} we observe a significant increase in the intensity noise observed on the cavity transmission compared to the laser directly after the amplifier. The increase in the RIN spectrum can be understood as converted from a combination of phase noise of the laser and vibrations of the cavity, which can be mitigated with further technical effort. The attribution of the heating to the RIN is quantitatively plausible, and also supported by the fact that we observe a 2 times shorter lifetime for ground state atoms in tweezers, for which the \SI{1036}{nm} light has a 4 times larger polarizability. Another potential hypothesis for the short lifetime is collisions with atoms from the reservoir. However, we observe that the lifetime of $^1S_0$ atoms remains unchanged regardless of the atomic density or flux in the ODT, suggesting that this is not a dominant contribution.

\end{document}